\documentclass[english,twocolumn,pra,eqsecnum, showpacs]{revtex4}
\usepackage[T1]{fontenc}
\usepackage[latin1]{inputenc}
\usepackage{amsmath}
\usepackage{graphicx}
\usepackage{amssymb}

\makeatletter

\makeatletter

\makeatletter

\makeatletter
\usepackage{epsf}

\makeatother

\makeatother

\makeatother

\usepackage{babel}
\makeatother

\begin{document}

\title{Multi-component strongly attractive Fermi gas: a color superconductor
in a one-dimensional harmonic trap}

\author{Xia-Ji Liu$^{1,2}$, Hui Hu$^{2}$, and Peter D. Drummond$^{1}$}

\affiliation{$^{1}$\ ARC Centre of Excellence for Quantum-Atom Optics, School
of Physical Sciences, University of Queensland, Brisbane, Queensland
4072, Australia \\
 $^{2}$\ Department of Physics, Renmin University of China, Beijing
100872, China}

\date{\today}

\begin{abstract}
Recent advances in ultra-cold atomic Fermi gases make it possible
to achieve a fermionic superfluid with multiple spin components. In
this context, any mean-field description is expected to fail, owing
to the presence of tightly bound clusters or molecules that consist
of more than two particles. Here we present a detailed study of a
strongly interacting multi-component Fermi gas in a highly elongated
or quasi-one-dimensional harmonic trap, which could be readily obtained
in experiment. By using the exact Bethe ansatz solution and a local
density approximation treatment of the harmonic trap, we investigate
the equation of state of the multi-component Fermi gas in both a homogeneous
and trapped environment, as well as the density profiles and low-energy
collective modes. The binding energy of multi-component bound clusters
is also given. We show that there is a peak in the collective mode
frequency at the critical density for a deconfining transition to
a many-body state that is analogous to the quark color superconductor
state expected in neutron stars. 
\end{abstract}

\pacs{03.75.Ss, 05.30.Fk, 71.10.Pm, 74.20.Fg}

\maketitle

\section{Introduction}

Recent experimental progress achieved in trapped ultra-cold atomic
gases provides a great opportunity for exploring strongly interacting
many-body physics. Owing to the molecular Feshbach resonance \citep{FR},
the strength of the interactions between atoms in different hyper-fine
states (or species) can be arbitrarily changed from strong to weak
coupling in a well-controlled manner. Moreover, interactions can be
effectively tuned by using optical lattices with a varying tunneling
barrier \citep{lattice}. Consequently, a number of many-body models
in condensed matter physics and nuclear physics can be readily accessed
in the ultra-cold atomic gases. In the past few years, considerable
interest has been focused on trapped two-component Fermi gases close
to a broad Feshbach resonance. In particular, the crossover from Bardeen-Cooper-Schrieffer
(BCS) fermionic superfluidity of Cooper pairs to Bose-Einstein condensation
(BEC) of tightly bounded molecules has been explored in great detail
\citep{huinp,leggett,nsr,randeria,griffin,haussmann,hui04,hld,jila,mit04,duke,chin,ens,mit05,randy05,randy06}.

Multi-component Fermi gases with more than two species can be easily
trapped and manipulated as well. For this type of Fermi gas, bound
{\em multi-body} clusters are expected to appear above a critical
interaction strength \citep{luu}, in addition to the formation of
Cooper pairs due to two-body {}``pairing correlations''. As is well
known, a landmark theoretical result in quantum physics is Efimov's
prediction of a universal set of bound trimer states appearing for
three identical bosons with a resonant two-body interaction \citep{efimov}.
The existence of such an Efimov resonance greatly changes the properties
of dilute Bose gases, as observed recently in an ultracold gas of
cesium atoms \citep{grimm}. Similarly, multi-body clusters are of
fundamental importance to fermionic superfluidity in multi-component
Fermi gases. While Cooper pairs are dominant in the weak coupling
limit, an exotic superfluid state with bound multi-body clusters should
emerge in the strong coupling regime. In between, a quantum phase
transition is then anticipated to take place (as specified for the
three component case in Ref. \citep{hofstetter3}), in contrast to
the smooth crossover observed in the two-component case \citep{huinp}.

This issue is relevant to outstanding problems in nuclear and particle
physics. The quark model of nuclear matter at low density describes
nucleons as three fermion clusters: tri-quark bound states. At sufficiently
high density and pressure, it is conjectured that a phase-transition
occurs to a deconfined color superfluid phase of quark matter \citep{BailinLove84,Alford}.
This is believed to occur in the interior of neutron stars and possibly
in heavy-ion collisions. While quark matter has many other features,
it is interesting to find that a physically accessible system of interacting
ultra-cold fermions is also expected to display these features of
multi-body fermion clusters and quantum phase-transitions. Ultra-cold
atoms could provide a means to test these theoretical predictions.
If confirmed, this would lend support to current ideas in particle
theory and astrophysics.

The description of multi-body bound states is beyond the generally
adopted mean-field framework, which treats competition between two-body
correlations among different atom pairs. This, however, is of importance
only in the weak coupling limit. For strong coupling, the use of numerical
Monte Carlo techniques is hampered by the fermionic sign problem \citep{Allton}.
It is therefore of great importance to have an analytically soluble
model of multi-component Fermi gases, and to study both the weakly
and strongly interacting regimes on an equal footing. An exact analysis
is provided in this paper for the case of multi-component Fermi gases
in one dimension (1D), where multi-body bound clusters are always
present, regardless of the interaction strength. Although the one
dimensional analysis gives a smooth crossover rather than a true phase
transition, we believe that considerable insight may be obtained for
a three dimensional gas as well.

As well as being exactly soluble, the 1D problem has great experimental
relevance. A quantum degenerate trapped two-component Fermi gas in
quasi 1D has been demonstrated recently by loading an ultra-cold Fermi
gas into a two-dimensional optical lattice \citep{esslinger} of trapping
`tubes'. In this configuration, the radial motion of atoms perpendicular
to each tube is frozen to zero-point oscillations due to tight transverse
confinement, while axial motion is only weakly confined. One then
obtains an array of effective 1D systems, each in an axial harmonic
trap. The manipulation of more than two species in 1D is within reach
of present-day technology, and is likely to be achieved soon in experiments.

A typical example of these developments is lithium gas \citep{a3dB},
which has favorable collisional properties among its lowest three
hyperfine spin states, denoted $\left|1\right\rangle $, $\left|2\right\rangle $
and $\left|3\right\rangle $, respectively. A recent accurate measurement
of the scattering lengths between these hyperfine states shows that
the background interactions are anomalously large \citep{a3dB}, with
background scattering lengths about $-1500a_{0}$, where $a_{0}$
($=0.0529177$ nm) is the Bohr radius. There are also three broad
$s$-wave Feshbach resonances located at the positions $B=834,811,$
and $690$ Gauss for ($1,2$), ($2,3$), and ($3,1$) channels, respectively.
These peculiar collisional properties are useful to cool the gas down
to the quantum degenerate regime. Ideally, one expects experimentally
accessible lowest temperatures for this three-state mixture to be
in the same range as for two-component ensembles, \textit{i.e.}, $T\simeq0.05T_{F}$,
where $T_{F}$ is the Fermi temperature of an ideal Fermi gas. Thus,
a novel fermionic multi-component superfluid may be anticipated. For
this reason, three-component lithium gas has attracted a great deal
of theoretical interest, including analysis of mean-field states \citep{hofstetter1,hofstetter2,torma1,lianyi,cherng}
as well as phase diagrams \citep{bedaque,zhai,sedrakian,torma2,hofstetter3,Capponi,Lecheminant,Wu}.

Here, we report on properties of a multi-component Fermi gas in 1D.
Firstly, using the exact Bethe ansatz solution \citep{lieb,yang,gaudin,takahashi1},
we investigate the exact ground state of a homogeneous gas with attractive
interactions at zero temperature. To make contact with experiments,
we then consider an inhomogeneous Fermi gas under harmonic confinement,
within the framework of the local density approximation (LDA). The
equation of state of the system in both the uniform and trapped case
are investigated in detail. Particular attention is drawn to the density
profiles and low-lying collective modes of the trapped cloud, which
are readily measurable in experiment. We show that the gas becomes
more attractive as the number of species increases, demonstrating
the strongly interacting nature of multi-body bound clusters.

A 1D multi-component Fermi gas in an optical lattice was considered
recently by Capponi and co-workers \citep{Capponi}, using both the
analytic bosonization approach and the numerical density matrix renormalization
group method. This lattice version of the 1D Fermi gas resembles the
system we consider here. In particular, it also features a molecular
superfluid phase in the low density limit with a strongly attractive
interaction \citep{Capponi}.

The paper is organized as follows. In the following section, we outline
the theoretical model for a 1D multi-component Fermi gas. Of particular
relevance for an experimental realization is our calculation of the
effective 1D coupling constant for the three-component lithium gases.
In Sec. III, we present the exact Bethe ansatz solution and discuss
the equation of state and the sound velocity of a uniform system at
zero temperature. In Sec. IV, using the LDA method we investigate
the density profile and the equation of state in the trapped environment.
Also, we describe the dynamics of trapped gases in terms of 1D hydrodynamic
equations and develop a novel algorithm to solve these equations.
The behavior of low-lying collective modes is then obtained and discussed.
We end with some concluding remarks in Sec. V. An appendix is used
to outline the details of the algorithm used in solving the 1D hydrodynamic
equations.

\section{Models}

A quasi-1D multi-component Fermi gas in a highly elongated trap can
be formed using a two-dimensional optical lattice \citep{esslinger}.
By suitably tuning the lattice depth, the anisotropy aspect ratio
$\lambda=\omega_{z}/\omega_{\rho}$ of two harmonic frequencies can
become extremely small. This means that the Fermi energy associated
with the longitudinal motion of the atoms is much smaller than the
energy level separation along the transverse direction, \textit{i.e.},
$N\hbar\omega_{z}\ll\hbar\omega_{\rho}$ and $k_{B}T\ll\hbar\omega_{\rho}$,
where $N$ is the total number of atoms. In this limit, the transverse
motion will be essentially frozen out, and one ends up with a quasi-one
dimensional system.

\subsection{Interaction Hamiltonian}

We study a gas with pseudo-spin $S=\left(\kappa-1\right)/2$, where
$\kappa$ ($\geq2$) is the number of components. From now we shall
assume that the fermions in different spin states attract each other
via the {\em same} short-range potential $g_{1D}\delta(x)$. Denoting
the mass of each fermion as $m$, with a total fermion number $N=\sum_{l=1}^{\kappa}N_{l}$
(where $N_{l}$ is the number of fermions with pseudo-spin projection
$l$) the first quantized Hamiltonian for the system is therefore
\begin{equation}
{\cal H}={\cal H}_{0}+\sum_{i=1}^{N}\frac{1}{2}m\omega^{2}x_{i}^{2}.\end{equation}
 Here \begin{equation}
{\cal H}_{0}=-\frac{\hbar^{2}}{2m}\sum_{i=1}^{N}\frac{\partial^{2}}{\partial x_{i}^{2}}+g_{1D}\sum_{i<j}\delta(x_{i}-x_{j})\end{equation}
 represents the part of Hamiltonian in free space without the trapping
potential $m\omega^{2}x^{2}/2$, while $\omega=\omega_{z}$ is an
oscillation frequency in the axial direction. There is an inter-particle
attraction between any two fermions with different quantum numbers.

In an elongated trap, the 1D effective coupling constant $g_{1D}$
is related to the 3D scattering length $a_{3D}$. It is shown by Bergeman
\textit{et al.} \citep{bergeman,astrakharchik} that \begin{equation}
g_{1D}=\frac{2\hbar^{2}a_{3D}}{ma_{\rho}^{2}}\frac{1}{\left(1-Aa_{3D}/a_{\rho}\right)},\end{equation}
 where $a_{\rho}=\sqrt{\hbar/(m\omega_{\rho})}$ is the characteristic
oscillator length in the transverse axis. The constant $A=-\zeta(1/2)/\sqrt{2}\simeq1.0326$
is responsible for a confinement induced Feshbach resonance \citep{footnote},
which changes the scattering properties dramatically when the 3D scattering
length is comparable to the transverse oscillator length. It is convenient
to express $g_{1D}$ in terms of an effective1D scattering length,
$g_{1D}=-2\hbar^{2}/\left(ma_{1D}\right)$, where \begin{equation}
a_{1D}=-\frac{a_{\rho}^{2}}{a_{3D}}\left(1-A\frac{a_{3D}}{a_{\rho}}\right)>0.\label{a1d}\end{equation}
 Note that in this definition of the 1D scattering length, our sign
convention is opposite to the 3D case.

In the homogeneous case, we measure the interactions by a dimensionless
coupling constant $\gamma$, which is the ratio of the interaction
energy density $e_{int}$ to the kinetic energy density $e_{kin}$
\citep{lieb}. In the weak coupling, $e_{int}\sim g_{1D}n$ and $e_{kin}\sim\hbar^{2}k^{2}/(2m)\sim\hbar^{2}n^{2}/m$,
where $n$ is the total linear density of the gas. Therefore, one
finds \begin{equation}
\gamma=-\frac{mg_{1D}}{\hbar^{2}n}=\frac{2}{na_{1D}}.\end{equation}
 Thus, $\gamma\ll1$ corresponds to the weakly interacting limit,
while the strong coupling regime is realized when $\gamma\gg1$.

\subsection{Cluster states}

In the case where all the fermions are distinct - which is only possible
if the number of fermions is less than or equal to $\kappa$ - the
spatial wave-function can be completely symmetric. This allows one
to construct eigenstates with identical symmetry to the exact solutions
already known for a one-dimensional Bose gas with attractive interactions
\citep{McGuire}. This exceptionally simple limiting case gives useful
physical insight into the multi-particle clusters. These will be an
essential feature in the physical properties of more general solutions.
Accordingly, we may consider as a trial solution the localized quantum
soliton state with wavefunction:

\begin{equation}
\Psi\left(x_{1}\ldots x_{\kappa}\right)=\exp\left\{ -c\sum_{i>j}\left|x_{i}-x_{j}\right|\right\} \end{equation}

On calculating the effect of the many-body Hamiltonian, we find that:\begin{equation}
H_{0}\Psi=E_{\kappa}\Psi+\left[\frac{2\hbar^{2}c}{m}+g_{1D}\right]\left\{ \sum_{i>j}\delta\left(x_{i}-x_{j}\right)\right\} \Psi\,,\end{equation}
 where the energy $E_{\kappa}$ is: \begin{equation}
E_{\kappa}=\frac{-\hbar^{2}c^{2}\kappa(\kappa^{2}-1)}{6m}\,.\end{equation}
 This symmetric state with an asymptotic exponentially decaying wavefunction
in each coordinate direction is the fundamental bound cluster. The
requirement for this to be an eigenstate is simply:

\begin{equation}
c=\frac{-mg_{1D}}{2\hbar^{2}}=\frac{1}{a_{1D}}>0\,.\end{equation}
 In terms of this characteristic length-scale of $a_{1D}=1/c$, one
finds that the fundamental cluster binding energy can be written as:\begin{equation}
\epsilon_{B}=\frac{\hbar^{2}\kappa(\kappa^{2}-1)}{6ma_{1D}^{2}}\,.\label{eq:clusterbinding}\end{equation}

These bound clusters can have either a fermionic or bosonic character,
depending on whether $\kappa$ is odd or even. The binding energy
scales quadratically with the Hamiltonian coupling, and cubically
with the number of bound fermions, $\kappa.$ Clusters are localized
relative to the center of mass, with a characteristic length scale
of $a_{1D}$. One may reasonably expect some kind of physical transition
to occur when the linear density exceeds $1/a_{1D}$, since at high
density the Pauli exclusion principle will not allow independent clusters
to form.

Thus, we can expect this type of symmetric bound state to predominate
at low density, with a transition to a radically different behaviour
at high density. This transition due to the Pauli exclusion principle
is a unique feature of a 1D attractive Fermi gas, and does not occur
in a 1D attractive Bose gas. By contrast, in an attractive Bose gas,
clusters or quantum solitons can form with arbitrary particle number.
They have been already observed with up to for $N=10^{7}$ for photons
in optical fibers \citep{DrummondNature}, and $N=10^{3}$ in the
case of ultra-cold atoms \citep{Strecker}. However, clearly this
is not to be expected in the case of fermions, where the spin multiplicity
limits the size of this type of symmetric cluster.

For $\kappa=3$, here is a close analogy between the symmetry properties
of these bound clusters and the color symmetry properties of quark
models in particle physics. In the case of quark matter, free nucleons
are three-quark bound states. However, at high density, it is conjectured
that there is a quantum phase-transition to a deconfined color superconductor
state \citep{BailinLove84,Alford}. This is expected physically at
the core of massive neutron stars or in heavy-ion collisions. Although
we are interested mainly in the one-dimensional case, where mostly
a crossover would be expected due to dimensionality, we will show
that a similar type of deconfining {}``transition'' occurs here
also.

\subsection{Harmonic Trap}

In the presence of a harmonic trap, we may characterize the interactions
using the dimensionless coupling constant at the trap center $\gamma_{0}$.
For an ideal Fermi gas with equal spin population in each component,
the total linear density is \begin{equation}
n_{ideal}\left(x\right)=n_{TF,\kappa}^{0}\left[1-\frac{x^{2}}{\left(x_{TF,\kappa}^{0}\right)^{2}}\right]^{1/2},\end{equation}
 in the large-$N$ Thomas-Fermi (TF) limit, where \begin{eqnarray}
n_{TF,\kappa}^{0} & = & \frac{\left(2N\kappa\right)^{1/2}}{\pi}a_{ho}^{-1},\\
x_{TF,\kappa}^{0} & = & \left(\frac{2N}{\kappa}\right)^{1/2}a_{ho},\end{eqnarray}
 are the center linear density and the TF radius respectively . Here
$a_{ho}=\sqrt{\hbar/(m\omega)}$ is the characteristic oscillator
length in the axial direction. We therefore find that: \begin{equation}
\gamma_{0}=\left(\frac{2}{\kappa}\right)^{1/2}\pi\left[\frac{1}{N^{1/2}}\left(\frac{a_{ho}}{a_{1D}}\right)\right].\end{equation}
 To remove the dependence on the number of components $\kappa$, we
define a dimensionless parameter \begin{equation}
\delta=\left[N\frac{a_{1D}^{2}}{a_{ho}^{2}}\right]^{-1/2}\label{delta}\end{equation}
 to describe the interactions. Note that the parameter $\delta$ depends
inversely on the total number of particles. Hence, somewhat counter-intuitively,
the gas becomes increasingly non-ideal with a decreasing number of
atoms.

%
\begin{figure}
\begin{centering}
\includegraphics[clip,width=0.45\textwidth]{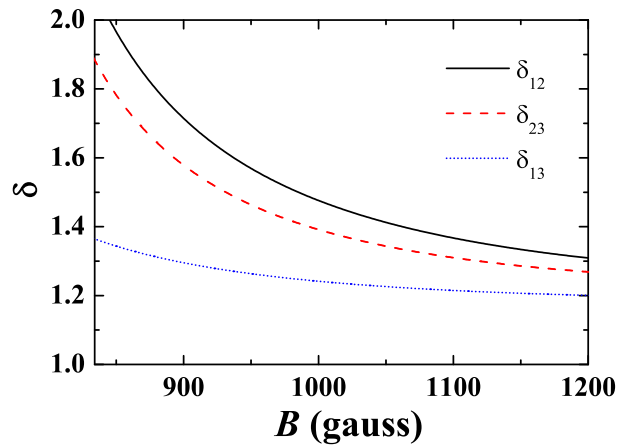} 
\par\end{centering}

\caption{(Color online) One-dimensional interaction parameters for a three-component
lithium gas in a two-dimensional optical lattice above the Feshbach
resonances. Above a magnetic field $B=1000$ Gauss, the difference
of interaction parameters between different channels becomes very
small, \textit{i.e.}, $(\delta_{12}-\delta_{23})/\delta_{12}<0.06$
and $(\delta_{12}-\delta_{13})/\delta_{12}<0.16$. As a result, a
three-component lithium gas can be well described by our exactly soluble
model of 1D Fermi gases with a symmetric interaction.}

\label{fig1} 
\end{figure}

\subsection{Parameter values}

For experimental relevance, we now estimate interaction parameters
for the on-going experiments on 1D lithium gases. Consider a gas of
$^{6}$Li atoms in 2D optical lattice with a typical lattice spacing
periodicity $d=532$ nm. The transverse oscillator length $a_{\rho}$
is related to $d$ via $a_{\rho}=d/(\pi s^{1/4})$ \citep{zwerger},
where $s$ is the ratio of the lattice depth to the recoil energy.
Taking $s=10$, this equation yields $a_{\rho}\simeq95$ nm. Empirically,
the 3D scattering length of $^{6}$Li gases at these broad resonances
is given by \begin{equation}
a_{3d}=a_{b}[1+W/(B-B_{0})][1+\alpha(B-B_{0})]\,.\end{equation}
 Detailed values of the background scattering length $a_{b}$, resonance
position $B_{0}$, resonance width $W$, and leading-order correction
$\alpha$ have been measured precisely by Bartenstein \textit{et al.}
\citep{a3dB} for all three channels. As an example, we take a trapping
frequency of $\omega=\omega_{z}=2\pi\times400$ Hz, which gives rise
to $a_{ho}=2052$ nm. The number of atoms in one tube of the lattice
can be approximately of order $N\sim100$. Given these parameters,
we use Eqs. (\ref{a1d}) and (\ref{delta}) to calculate $\delta$.
The estimated dimensionless coupling constants are shown in Fig. (\ref{fig1}).
Here, we focus on the BCS side of the resonances, and thus take a
magnetic field $B>834$ Gauss. We find that $\delta\sim O(1)$, \textit{i.e.},
the gas is in an intermediate interacting regime. The difference in
the interactions between different channels turns out be small above
1000 Gauss. This justifies our choice of the same contact interaction
potential between different hyperfine spin states.

\section{Homogeneous multi-component Fermi gases}

We first consider a uniform multi-component Fermi gas in one dimension
with symmetric inter-component interactions. In this case, the model
is exactly soluble via the Bethe ansatz \citep{lieb,yang,gaudin,takahashi1,astrakharchik,ldh1d,hld1d,guan1,guan2,guan3,takahashi2,Schlottmann}.
For simplicity, we assume that each component has the same number
of particles, \textit{i.e.}, $N_{l}\equiv N/\kappa$ ($l=1,2,\cdots,$
$\kappa=2S+1$).

\subsection{Ground State}

In the ground state, the particles partition into groups of $\kappa$
fermions. In each group, the fermions all have different quantum numbers,
and are bound together to form a $\kappa$-body cluster. Introducing
a linear number density, $n=N/L$, where $L$ is the size of the system,
the ground state energy $E_{\hom}$ in the thermodynamic limit is
given by \citep{takahashi1,takahashi2}, \begin{equation}
\frac{E_{\hom}}{L}=\frac{\hbar^{2}}{2m}\int\limits _{-Q}^{Q}d\Lambda\left(\kappa\Lambda^{2}-\frac{\kappa\left(\kappa^{2}-1\right)}{3}c^{2}\right)\rho\left(\Lambda\right),\end{equation}
 where the coupling $c=n\gamma/2=1/a_{1D}$ and $\rho(\Lambda)$ is
the quasi-momentum distribution of $\kappa$-body clusters with a
cut-off rapidity $Q$. The quasi-momentum distribution is determined
by an integral equation \citep{takahashi1,takahashi2}, \begin{equation}
\rho\left(\Lambda\right)=\frac{\kappa}{2\pi}-\sum\limits _{l=1}^{\kappa-1}\int\limits _{-Q}^{Q}d\Lambda^{\prime}\frac{2lc\rho\left(\Lambda^{\prime}\right)}{\left(2lc\right)^{2}+\left(\Lambda-\Lambda^{\prime}\right)^{2}},\end{equation}
 and is normalized according to \begin{equation}
n=\kappa\int_{-Q}^{Q}d\Lambda\rho(\Lambda),\end{equation}
 which fix the value of the cut-off rapidity. The last term in $E_{\hom}$
is simply the contribution from $\kappa$-body bound states and is
equal to $-(n/\kappa)\epsilon_{\kappa b}$, with binding energy identical
to that given in the single-cluster result of Eq (\ref{eq:clusterbinding}):
\begin{equation}
\epsilon_{\kappa b}\equiv\frac{\hbar^{2}}{2m}\frac{\kappa\left(\kappa^{2}-1\right)}{3}c^{2}=\frac{\kappa\left(\kappa^{2}-1\right)}{6}\frac{\hbar^{2}}{ma_{1D}^{2}}.\end{equation}
 We note that the binding energy of multi-component clusters increases
rapidly with an increasing number of species $\kappa$. In particular,
when $\kappa\geq3$ it is larger than the pairing energy of $\kappa\left(\kappa-1\right)/2$
Cooper pairs. In other words, if the binding energy was solely due
to Cooper pairing, one would expect $\epsilon_{CP}=\kappa\left(\kappa-1\right)\epsilon_{2b}/2$,
where $\epsilon_{2b}=\hbar^{2}/ma_{1D}^{2}$ is the two-body binding
energy. The increase above this level is due to $\kappa$-body correlations:
$\kappa$ particles interact more strongly with each in a cluster
than as isolated pairs of particles.

\subsection{Sound velocity}

Once the ground state energy is obtained, we calculate the chemical
potential $\mu_{\hom}=\partial E_{\hom}/\partial N$ and the corresponding
sound velocity $c_{\hom}=\sqrt{n(\partial\mu_{\hom}/\partial n)/m}$.
The sound velocity will be utilized later for predicting measurable
collective mode frequencies. For numerical purposes, it is convenient
to rewrite these in a dimensionless form that depends on the dimensionless
coupling constant $\gamma$ only, \begin{eqnarray}
\frac{E_{\hom}}{L} & \equiv & \frac{\hbar^{2}n^{3}}{2m}\left[e\left(\gamma\right)-\frac{\left(\kappa^{2}-1\right)}{12}\gamma^{2}\right],\\
\mu_{\hom} & \equiv & \frac{\hbar^{2}n^{2}}{2m}\left[\mu\left(\gamma\right)-\frac{\left(\kappa^{2}-1\right)}{12}\gamma^{2}\right],\\
c_{\hom} & \equiv & \frac{\hbar n}{m}c\left(\gamma\right).\end{eqnarray}
 These are related by, \begin{eqnarray}
\mu\left(\gamma\right) & = & 3e\left(\gamma\right)-\gamma\frac{\partial e\left(\gamma\right)}{\partial\gamma},\\
c\left(\gamma\right) & = & \mu\left(\gamma\right)-\frac{\gamma}{2}\frac{\partial\mu\left(\gamma\right)}{\partial\gamma}.\end{eqnarray}
 It is easy to see that for an ideal multi-component Fermi gas, \begin{eqnarray}
\frac{E_{\hom}^{ideal}}{L} & = & \frac{\hbar^{2}n^{3}}{2m}\left(\frac{\pi^{2}}{3\kappa^{2}}\right),\\
\mu_{\hom}^{ideal} & = & \frac{\hbar^{2}n^{2}}{2m}\left(\frac{\pi^{2}}{\kappa^{2}}\right),\\
c_{\hom}^{ideal} & = & \frac{\hbar n}{m}\left(\frac{\pi}{\kappa}\right).\end{eqnarray}
 Therefore, as units of energy and sound velocity, we define a Fermi
energy $\varepsilon_{F,\kappa}\equiv[\hbar^{2}n^{2}/(2m)](\pi^{2}/\kappa^{2})$
and a Fermi velocity $\nu_{F,\kappa}\equiv[\hbar n/m](\pi/\kappa).$

\subsection{Numerical solutions}

The integral equation for the quasi-momentum distribution has to be
solved numerically. This was partly carried out by Schlottmann in
the $\kappa=4$ case, but for restricted values of the linear density
$n$ and coupling $c$ \cite{Schlottmann}. Here, for completeness,
we solve the integral equation for a general coupling constant $\gamma$.
The asymptotic behavior in the weak and strong coupling limits, not
explored in literature so far, will be discussed in detail.

To make the equation dimensionless, let us change variables as follows
\citep{lieb}: \begin{equation}
\Lambda\equiv Qx;\quad2c\equiv Q\lambda;\quad\rho\left(\Lambda\right)=g(x)\,.\end{equation}
 In terms of the new variables the quasi-momentum distribution, normalization
condition, and ground state energy become, respectively, \begin{eqnarray}
g\left(x\right) & = & \frac{\kappa}{2\pi}-\sum\limits _{\ell=1}^{\kappa-1}\frac{\ell}{\pi}\int\limits _{-1}^{+1}dx^{\prime}\frac{\lambda g\left(x^{\prime}\right)}{\left(\ell\lambda\right)^{2}+\left(x-x^{\prime}\right)^{2}},\nonumber \\
\lambda & = & \kappa\gamma\int\limits _{-1}^{+1}dxg\left(x\right),\nonumber \\
e\left(\gamma\right) & = & \frac{\gamma^{3}}{\lambda^{3}}\kappa\int\limits _{-1}^{+1}dxx^{2}g\left(x\right).\label{eg}\end{eqnarray}
 To obtain better numerical accuracy for the chemical potential, it
is useful to calculate the derivative of $e\left(\gamma\right)$ directly.
With this goal, we introduce $\lambda_{d}=d\lambda/d\gamma$ and $g_{d}(x)=dg(x)/d\gamma$,
which satisfy the coupled equations \begin{eqnarray}
g_{d}\left(x\right) & = & -\sum\limits _{\ell=1}^{\kappa-1}\frac{\ell}{\pi}\int\limits _{-1}^{+1}dx^{\prime}\left\{ \frac{\lambda g_{d}\left(x^{\prime}\right)}{\left(\ell\lambda\right)^{2}+\left(x-x^{\prime}\right)^{2}}\right.\nonumber \\
 &  & -\left.\frac{\lambda_{d}g\left(x\right)\left[-\left(\ell\lambda\right)^{2}+\left(x-x^{\prime}\right)^{2}\right]}{\left[\left(\ell\lambda\right)^{2}+\left(x-x^{\prime}\right)^{2}\right]^{2}}\right\} ,\nonumber \\
\lambda_{d} & = & \frac{\lambda}{\gamma}+\kappa\gamma\int\limits _{-1}^{+1}dxg_{d}\left(x\right).\label{gdx-norm}\end{eqnarray}
 The derivative of $e\left(\gamma\right)$ is then obtained from \begin{equation}
\frac{de}{d\gamma}=\frac{\gamma^{3}\kappa}{\lambda^{3}}\int\limits _{-1}^{+1}dxx^{2}\left[\left(\frac{3}{\gamma}-\frac{3\lambda_{d}}{\lambda}\right)g_{d}\left(x\right)+g\left(x\right)\right].\label{eg-deri}\end{equation}
 Numerically, the two set of integral equations, Eqs. (\ref{eg})
and Eqs. (\ref{gdx-norm}), have been solved by decomposing the integrals
on a grid with $M=1024$ points $\{x_{i};x_{i}\in\left[-1,+1\right]\}$.
For $g(x)$, we start from a set of trial distributions $g^{(0)}(x_{i})$,
with corresponding parameters of $\lambda^{(0)}$. Using the standard
method for the integrals \citep{lieb}, we obtain a new distribution
$g(x_{i})$, and update $\lambda$ accordingly. We iterate this procedure
until $g(x_{i})$ agrees with the previous distribution within a certain
tolerance, and finally calculate the energy function $e(\gamma)$
using Eq. (\ref{eg}). The integral equation of $g_{d}(x)$ can be
solved in the same manner, and finally Eq. (\ref{eg-deri}) gives
the derivative of the energy function $de/d\gamma$. We find that
these iterative procedures for solving the integral equations are
very stable. To obtain the sound velocity, the derivative of the chemical
potential has been calculated accurately as a numerical derivative.

%
\begin{figure}
\begin{centering}
\includegraphics[clip,width=0.45\textwidth]{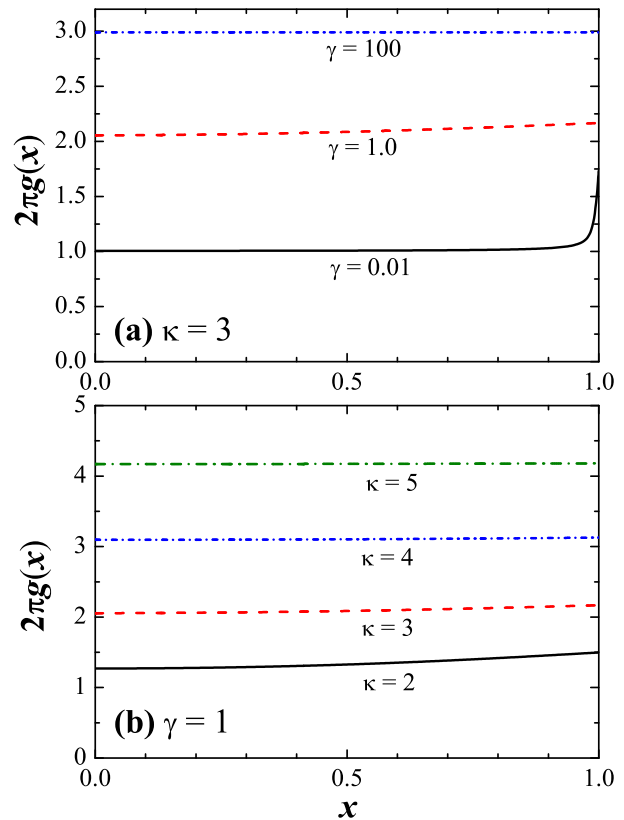} 
\par\end{centering}

\caption{(Color online) Quasi-momentum distributions as a function of the dimensionless
coupling constant (panel a) or as a function of the number of components
(panel b). The distribution approaches to $\kappa/2\pi$ for a large
interaction strength. While in the weak coupling limit, it goes to
$1/(2\pi)$, but with a sharp increase at the boundary of $x=\pm1$.}

\label{fig2} 
\end{figure}

For an illustrative purpose, we show in Fig. (\ref{fig2}) the quasi-momentum
distribution function $g(x)$ as a function of the coupling constant
(Fig. \ref{fig2}a) or as a function of the number of components (Fig.
\ref{fig2}b). As $g(x)$ is an even function, we plot only the part
with a positive $x$. For a large interaction strength, its approaches
$\kappa/2\pi$, while in the weak coupling limit, it reduces to $1/(2\pi)$.
Below we discuss this limiting behavior in detail.

\subsection{Strong coupling limit}

\begin{figure}
\includegraphics[width=0.45\textwidth]{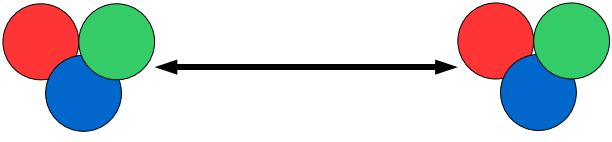}

\caption{Schematic diagram of interacting bound clusters for $\kappa=3$, in
the low density limit.}

\label{fig:cluster} 
\end{figure}

For a strongly interacting or equivalently, a low density gas, in
which the dimensionless coupling constant $\gamma\gg1$, the value
of $\lambda$ in Eq (\ref{gdx-norm}) is extremely large. Thus, the
integral kernel $l\lambda/\pi/[(l\lambda)^{2}+(x-x^{\prime})^{2}]$
becomes essentially a constant, $1/(\pi\ell\lambda)$. In addition,
the quasi-momentum distribution function $g(x)\simeq g_{0}$. Then,
the integral equation reduces to \begin{equation}
g_{0}=\frac{\kappa}{2\pi}-\sum\limits _{\ell=1}^{\kappa-1}\left(\frac{1}{\ell}\right)\frac{2}{\pi\lambda}g_{0}.\end{equation}
 At the same time, $\lambda=2\kappa\gamma g_{0}$. Denoting $S_{\kappa}=(1/\kappa^{2})\sum_{\ell=1}^{\kappa-1}\ell^{-1}$,
we find that up to the order $1/\gamma^{3}$, \begin{equation}
g\left(x\right)=\frac{\kappa}{2\pi}\left(1-\frac{2S_{\kappa}}{\gamma}\right)+O\left(\frac{1}{\gamma^{3}}\right),\quad\gamma\rightarrow\infty.\end{equation}
 Note that the factor $S_{\kappa}$ decreases rapidly as the number
of components increases, and goes like $S_{\kappa}\simeq\ln\kappa/\kappa^{2}$
when $\kappa\gg1$. It is then straightforward to obtain that, \begin{eqnarray}
e\left(\gamma\right) & = & \frac{\pi^{2}}{3\kappa^{4}}\left(1+\frac{4S_{\kappa}}{\gamma}+\frac{12S_{\kappa}^{2}}{\gamma^{2}}\right),\nonumber \\
\mu\left(\gamma\right) & = & \frac{\pi^{2}}{\kappa^{4}}\left(1+\frac{16S_{\kappa}}{3\gamma}+\frac{20S_{\kappa}^{2}}{\gamma^{2}}\right),\nonumber \\
c\left(\gamma\right) & = & \frac{\pi^{2}}{\kappa^{4}}\left(1+\frac{8S_{\kappa}}{\gamma}+\frac{40S_{\kappa}^{2}}{\gamma^{2}}\right),\end{eqnarray}
 and in dimensional form, \begin{eqnarray}
\frac{E_{\hom}}{N} & = & -\left(\frac{\epsilon_{\kappa b}}{\kappa}\right)+\frac{\hbar^{2}n^{2}}{2m}\frac{\pi^{2}}{3\kappa^{4}}\left[1+\frac{4S_{\kappa}}{\gamma}+\frac{12S_{\kappa}^{2}}{\gamma^{2}}\right],\nonumber \\
\mu_{\hom} & = & -\left(\frac{\epsilon_{\kappa b}}{\kappa}\right)+\frac{\hbar^{2}n^{2}}{2m}\frac{\pi^{2}}{\kappa^{4}}\left[1+\frac{16S_{\kappa}}{3\gamma}+\frac{20S_{\kappa}^{2}}{\gamma^{2}}\right],\nonumber \\
c_{\hom} & = & \frac{\hbar n}{m}\frac{\pi}{\kappa^{2}}\left[1+\frac{8S_{\kappa}}{\gamma}+\frac{40S_{\kappa}^{2}}{\gamma^{2}}\right]^{1/2}.\label{c-strong}\end{eqnarray}
 The first term on the right hand side of the total energy and chemical
potential is simply the binding energy per particle of a 1D bound
cluster from Eq (\ref{eq:clusterbinding}), while the second term
arises from interactions between clusters. These are strongly repulsive
due to the Pauli exclusion principle between identical fermions. In
the infinite coupling constant limit, the system behaves like a spinless
Tonks-Girardeau gas of bound clusters with hard-core repulsive interactions.
This is shown schematically in Fig (\ref{fig:cluster}).

Each cluster has a density $n/\kappa$ and mass $\kappa m$ \citep{tg},
giving rise to a chemical potential $\hbar^{2}\pi^{2}(n/\kappa)^{2}/(2\kappa m)=[\hbar^{2}/2m](\pi^{2}/\kappa^{3})$.
This is exactly $\kappa$ times the second term in the chemical potential
$\mu_{\hom}$. It is worth emphasizing that the compressibility of
the system remains positive for $\gamma\rightarrow\infty$ as indicated
by the first term in the sound velocity $c_{\hom}$, $(\hbar n/m)(\pi/\kappa^{2})$.
This means that a 1D multi-component Fermi gas is mechanically stable,
even in the strongly attractive regime. In contrast, the mechanical
stability of a strongly interacting 3D multi-component gas with $\kappa\geq3$
is not known exactly. It may experience collapse for sufficient large
number of components, as suggested by Heiselberg \citep{heiselberg}.
The parity of the number of species $\kappa$ may also play an important
role on influencing the stability in this limit. For the stable, 1D
case treated here, this low-density regime is analogous to the regime
of isolated nucleons in QCD.

\subsection{Weak coupling limit}

The asymptotic behavior in the weak coupling limit is more subtle.
Numerical calculation in the small $\gamma$ limit suggests that $g\left(x\right)\rightarrow1/(2\pi)$
and $\lambda\sim\gamma\rightarrow0$. We then expand the quasi-momentum
distribution, \begin{equation}
g\left(x\right)=\frac{1}{2\pi}+f\left(x\right),\end{equation}
 where $f\left(x\right)=O(\gamma)\ll1$ satisfies, \begin{eqnarray}
f\left(x\right) & = & \frac{\kappa-1}{2\pi}-\sum\limits _{\ell=1}^{\kappa-1}\frac{\ell}{\pi}\int\limits _{-1}^{+1}dx^{\prime}\frac{\lambda}{\left(\ell\lambda\right)^{2}+\left(x-x^{\prime}\right)^{2}}\frac{1}{2\pi}\nonumber \\
 &  & -\sum\limits _{\ell=1}^{\kappa-1}\frac{\ell}{\pi}\int\limits _{-1}^{+1}dx^{\prime}\frac{\lambda}{\left(l\lambda\right)^{2}+\left(x-x^{\prime}\right)^{2}}f\left(x\right).\end{eqnarray}
 As $\lambda\rightarrow0$, to the leading order $\ell\lambda/\pi/[(\ell\lambda)^{2}+(x-x^{\prime})^{2}]\simeq\delta(x-x^{\prime})$.
Thus, \begin{equation}
f\left(x\right)=\frac{1}{2\pi\kappa}\left[\kappa-1-\sum\limits _{\ell=1}^{\kappa-1}\int\limits _{-1}^{+1}\frac{dx^{\prime}\left(\ell\lambda/\pi\right)}{\left(\ell\lambda\right)^{2}+\left(x-x^{\prime}\right)^{2}}\right].\end{equation}
 For small $\lambda$, the integral in $f\left(x\right)$ is well
approximated by, \begin{equation}
\frac{\ell}{\pi}\int\limits _{-1}^{+1}dx^{\prime}\frac{\lambda}{\left(\ell\lambda\right)^{2}+\left(x-x^{\prime}\right)^{2}}\simeq1-\frac{2}{\pi}\frac{\ell\lambda}{1-x^{2}}.\end{equation}
 We find then to the order of $\gamma$, \begin{equation}
f\left(x\right)=\frac{\kappa-1}{2\pi^{2}}\frac{\lambda}{1-x^{2}},\end{equation}
 which diverges at the boundary $x=\pm1$. It is straightforward to
show that, \begin{eqnarray}
\int\limits _{-1}^{+1}dxg\left(x\right) & = & \frac{1}{\pi}-\frac{\kappa-1}{2\pi^{2}}\lambda\ln\lambda+\cdots,\\
\int\limits _{-1}^{+1}dx\left(1-x^{2}\right)g\left(x\right) & = & \frac{2}{3\pi}+\frac{\kappa-1}{\pi^{2}}\lambda+\cdots.\end{eqnarray}

%
\begin{figure}
\begin{centering}
\includegraphics[clip,width=0.45\textwidth]{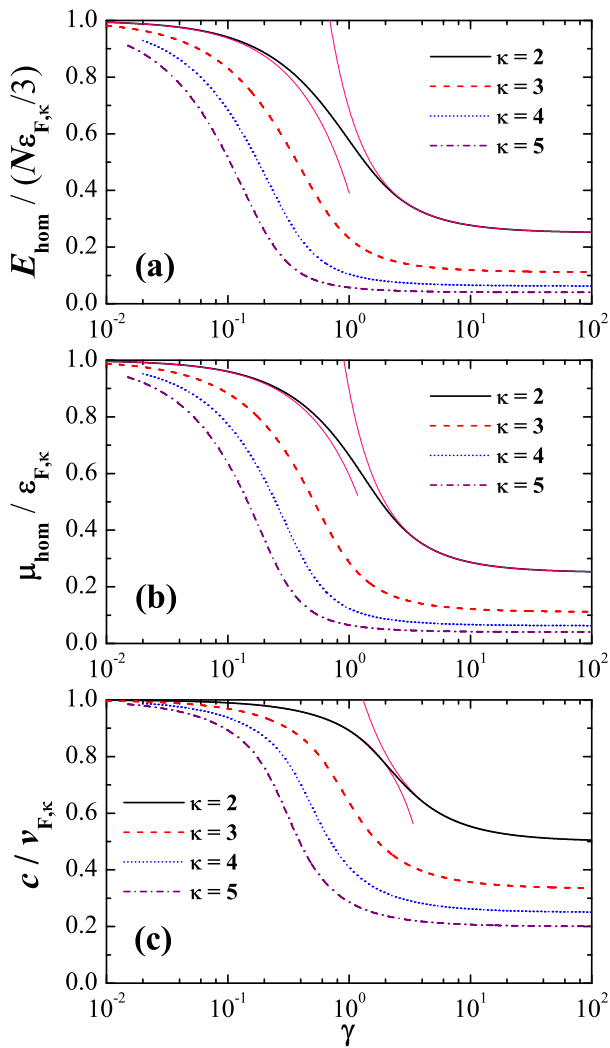} 
\par\end{centering}

\caption{(Color online) Dependence of the uniform ground state energy per particle
(a), the chemical potential (b), and the velocity of sound (c) on
the dimensionless coupling constant $\gamma$, at several number of
species as indicated. The energy and sound velocity are in units of
the Fermi energy $\varepsilon_{F,\kappa}\equiv[\hbar^{2}n^{2}/(2m)](\pi^{2}/\kappa^{2})$
and the Fermi velocity $\nu_{F,\kappa}\equiv[\hbar n/m](\pi/\kappa)$,
respectively. Thin solid lines are the analytic results in the two
limiting cases for $\kappa=2$, as described in Eq. (\ref{c-strong})
and Eq. (\ref{c-weak}).}

\label{fig4} 
\end{figure}

Substituting these results into the energy function $e(\gamma)$,
it is easy to obtain, \begin{equation}
e\left(\gamma\right)=\frac{\pi^{2}}{3\kappa^{2}}-\frac{\pi\left(\kappa-1\right)}{\kappa^{2}}\lambda-\frac{\left(\kappa-1\right)^{2}}{4\kappa^{2}}\lambda^{2}\ln^{2}\lambda.\end{equation}
 Recall that $\lambda=\kappa\gamma\int\nolimits _{-1}^{+1}g\left(x\right)\approx\kappa\gamma/\pi$.
The equation of state and the sound velocity are finally given by,
\begin{eqnarray}
\frac{E_{\hom}}{N} & = & \frac{\hbar^{2}n^{2}}{2m}\left[\frac{\pi^{2}}{3\kappa^{2}}-\frac{\left(\kappa-1\right)}{\kappa}\gamma-\frac{\left(\kappa-1\right)^{2}}{4\pi^{2}}\gamma^{2}\ln^{2}\gamma\right],\nonumber \\
\mu_{\hom} & = & \frac{\hbar^{2}n^{2}}{2m}\left[\frac{\pi^{2}}{\kappa^{2}}-2\frac{\left(\kappa-1\right)}{\kappa}\gamma-\frac{\left(\kappa-1\right)^{2}}{4\pi^{2}}\gamma^{2}\ln^{2}\gamma\right],\nonumber \\
c_{\hom} & = & \frac{\hbar n}{m}\frac{\pi}{\kappa}\left[1-\frac{\kappa\left(\kappa-1\right)}{\pi^{2}}\gamma\right]^{1/2},\label{c-weak}\end{eqnarray}
 where the first terms on the right hand side are identical to an
ideal (non-interacting) multi-component Fermi gas, as one might expect.
Note that the binding energy of bound clusters is of second order
in $\gamma$, and therefore is not included in the above expressions.
Two-body correlations are dominant in the weakly interacting limit,
and give rise to the mean-field Hartree-Fock attractive corrections
in the second term on the right hand side. The non-perturbative terms
of order $\gamma^{2}\ln^{2}\gamma$ are beyond mean-field theory.
This regime is analogous to the color-superconductor regime expected
in quark matter.

\subsection{Numerical results}

In Fig. (\ref{fig4}), we give the equation of state and the sound
velocity as a function of the dimensionless coupling constant $\gamma$,
obtained by numerically solving the integral equations. The ground
state energy per particle and the chemical potential are measured
in units of one-third of the Fermi energy and the Fermi energy, respectively,
while the sound velocity is in units of the Fermi velocity. We consider
a number of components ranging up to $\kappa=5$ to show the overall
trend.

Starting from the ideal gas results, the thermodynamic and dynamic
quantities decrease with increasing coupling constant, and finally
saturate to the Tonks-Girardeau (repulsive) gas limit, as already
anticipated. The rate of decrease is much faster as the number of
components $\kappa$ increases. This implies that the gas becomes
more attractive when the number of particles in the bound cluster
increases. We show also in the figure the asymptotic behavior in the
two limiting cases for $\kappa=2$ by thin solid lines. These fit
fairly well with the full numerical results, apart from a small intermediate
interaction region about $\gamma\sim1$.

It is worth noting that in the strongly interacting limit, the properties
of the uniform gas do not exhibit any (even-odd) parity dependence
on the number of species $\kappa$, because of Tonks-Girardeau fermionization,
as mentioned earlier. For a 1D multi-component gas in an optical lattice,
however, an entire different picture emerges \citep{Capponi}. Due
to the localization of the lattice in the strong attractive regime,
the system either becomes an ideal Fermi gas for odd $\kappa$, or
an ideal Bose gas for even $\kappa$, exhibiting a distinct parity
effect.

\section{Multi-component trapped Fermi gas}

To make quantitative contact with on-going experiments, it is crucial
to take into account the external harmonic trapping potential $V_{trap}(x)=m\omega^{2}x^{2}/2$
, which is necessary to prevent the fermions from escaping. For a
large number of fermions, which is likely to be $N\sim100$ experimentally,
an efficient way to take the trap into account is by using the local
density approximation (LDA). Together with the exact homogeneous equation
of state of a 1D multi-component Fermi gas, this gives an asymptotically
exact results as long as $N\gg1$.

The basic idea of the LDA is that an inhomogeneous gas of large size
can be treated locally as infinite matter with a local chemical potential.
We may then partition the cloud into many blocks, in each of which
the number of fermions is much greater than unity. Provided the variation
of the trap potential across each block is negligible compared with
the local Fermi energy, any interface effects may be safely neglected.
Thus, each block is un-correlated with the others. We note that in
1D the interface energy scales as $N^{-1}$ compared to the total
energy, and thus the LDA becomes valid provided $N\gg1$.

In detail, the LDA amounts to determining the chemical potential $\mu$
from the local equilibrium condition \citep{astrakharchik,ldh1d,hld1d},

\begin{equation}
\mu_{\hom}\left[n(x)\right]+\frac{1}{2}m\omega^{2}x^{2}=\mu_{g},\end{equation}
 under the normalization restriction, \begin{equation}
N=\int_{-x_{TF}}^{+x_{TF}}n\left(x\right)dx,\end{equation}
 where $n\left(x\right)$ is the total linear number density and is
nonzero inside a Thomas-Fermi radius $x_{TF}$. We have used the subscript
{}``$g$'' to distinguish the global chemical potential $\mu_{g}$
from the local chemical potential $\mu_{\hom}$. Rewriting $\mu_{\hom}$
into the dimensionless form $\mu[\gamma\left(x\right)]$, where $\gamma\left(x\right)=2/[n\left(x\right)a_{1D}]$,
we find that \begin{equation}
-\frac{\left(\kappa^{2}-1\right)\hbar^{2}}{6ma_{1D}^{2}}+\frac{\hbar^{2}n^{2}\left(x\right)}{2m}\mu\left[\gamma\left(x\right)\right]+\frac{1}{2}m\omega^{2}x^{2}=\mu_{g}.\end{equation}
 The first term on the left hand side is simply the binding energy,
and causes a constant shift to the chemical potential. To solve the
LDA equations, it is simplest to transform into a dimensionless form,
by defining \begin{eqnarray}
\widetilde{\mu}_{g} & = & \mu_{g}\frac{ma_{1D}^{2}}{\hbar^{2}}+\frac{\left(\kappa^{2}-1\right)}{6},\\
\widetilde{x} & = & \frac{a_{1D}x}{a_{ho}^{2}},\\
\widetilde{n} & = & na_{1D},\end{eqnarray}
 where the binding energy is now absorbed in the re-definition of
chemical potential. Thus, the local equilibrium condition becomes,
\begin{equation}
\frac{\widetilde{n}^{2}(\tilde{x})}{2}\mu\left[\gamma\left(\tilde{x}\right)\right]+\frac{\tilde{x}^{2}}{2}=\widetilde{\mu}_{g},\end{equation}
 where the dimensionless coupling constant now takes the form, $\gamma\left(\tilde{x}\right)=2/\widetilde{n}(\tilde{x})$.
Accordingly, the normalization condition is given by \begin{equation}
\int_{-\tilde{x}_{TF}}^{+\tilde{x}_{TF}}d\tilde{x}\tilde{n}(\tilde{x})=N\frac{a_{1D}^{2}}{a_{ho}^{2}}=\frac{1}{\delta^{2}},\end{equation}
 where $\delta$ is the interaction parameter for a trapped gas defined
earlier in Eq. (\ref{delta}). It is clear that the LDA equations
are controlled by a single parameter $\delta$: $\delta\ll1$ corresponds
to the weakly coupling limit, while $\delta\gg1$ corresponds to the
strongly interacting regime.

The numerical procedure of solving the LDA equations is straightforward.
For a given interaction parameter $\delta$, and initial guess for
$\widetilde{\mu}_{g}$, we invert the dimensionless local equilibrium
equations to find $\gamma(\tilde{x})$ and the linear density $\widetilde{n}\left(\tilde{x}\right)=2/\gamma\left(\tilde{x}\right)$.
The chemical potentials $\widetilde{\mu}_{g}$ are then adjusted to
enforce the number conservation, giving a better estimate for the
next iterative step. The iteration is continued until the number normalization
condition is satisfied within a certain range.

\subsection{Density profiles and the equation of state}

The asymptotic behavior of density profiles can be determined analytically
in the strong and weak coupling limits. In the strong interaction
regime of $\delta\gg1$, $\mu\left(\gamma\right)=(\pi^{2}/\kappa^{4})[1+16S_{\kappa}/(3\gamma)]$,
and the local equilibrium condition is given by, \begin{equation}
\frac{\tilde{n}^{2}\left(\tilde{x}\right)}{2}\frac{\pi^{2}}{\kappa^{4}}\left[1+\frac{16S_{\kappa}}{3\gamma\left(\tilde{x}\right)}\right]+\frac{\tilde{x}^{2}}{2}=\widetilde{\mu}_{g}.\end{equation}
 In the infinite coupling limit of Tonks-Girardeau gas, where $\gamma\left(\tilde{x}\right)\rightarrow\infty$
and therefore the second term in $\mu\left(\gamma\right)$ vanishes,
the density profile takes the form, \begin{equation}
\tilde{n}_{TG}\left(\tilde{x}\right)=\frac{\sqrt{2}\kappa}{\pi\delta}\left(1-\frac{\kappa^{2}\delta^{2}}{2}\tilde{x}^{2}\right)^{1/2},\end{equation}
 and the global chemical potential is, \begin{equation}
\widetilde{\mu}_{g}^{\left(0\right)}=\frac{1}{\kappa^{2}\delta^{2}}.\end{equation}
 The inclusion of the next order of $1/\gamma$ in $\mu\left(\gamma\right)$
leads to a density variation $\Delta\tilde{n}\left(\tilde{x}\right)$,
as well as a change in the chemical potential $\Delta\widetilde{\mu}_{g}$.
Linearizing the local equilibrium condition, we find that, \begin{equation}
\Delta\tilde{n}\left(\tilde{x}\right)=\frac{\kappa^{4}}{\pi^{2}}\frac{\Delta\widetilde{\mu}_{g}}{\tilde{n}_{TG}\left(\tilde{x}\right)}-\frac{8S_{\kappa}}{3}\left[\tilde{n}_{TG}\left(\tilde{x}\right)\right]^{2}.\end{equation}
 Number conservation $\int d\tilde{x}\Delta\tilde{n}(\tilde{x})=0$
yields, \begin{equation}
\Delta\widetilde{\mu}_{g}=\frac{64\sqrt{2}S_{\kappa}}{9\pi^{2}\kappa\delta^{3}}.\end{equation}
 Restoring the equations to dimensional form, the density profile
of a strongly interacting gas becomes, \begin{equation}
n\left(x\right)_{\delta\gg1}=n_{TG}\left(x\right)+\Delta n\left(x\right),\end{equation}
 where \begin{equation}
n_{TG}\left(x\right)=\sqrt{\kappa}n_{TF,\kappa}^{0}\left[1-\frac{\kappa x^{2}}{\left(x_{TF,\kappa}^{0}\right)^{2}}\right]^{1/2},\end{equation}
 is the profile of a spinless Tonks-Girardeau gas, and the density
variation, \begin{eqnarray}
\Delta n\left(x\right) & = & \frac{32\sqrt{2}\kappa^{3/2}S_{\kappa}}{9\pi^{2}\delta}n_{TF,\kappa}^{0}\left\{ \left[1-\frac{\kappa x^{2}}{\left(x_{TF,\kappa}^{0}\right)^{2}}\right]^{-1/2}\right.\nonumber \\
 &  & -\frac{3\pi}{4}\left.\left[1-\frac{\kappa x^{2}}{\left(x_{TF,\kappa}^{0}\right)^{2}}\right]\right\} .\end{eqnarray}

%
\begin{figure}
\begin{centering}
\includegraphics[clip,width=0.45\textwidth]{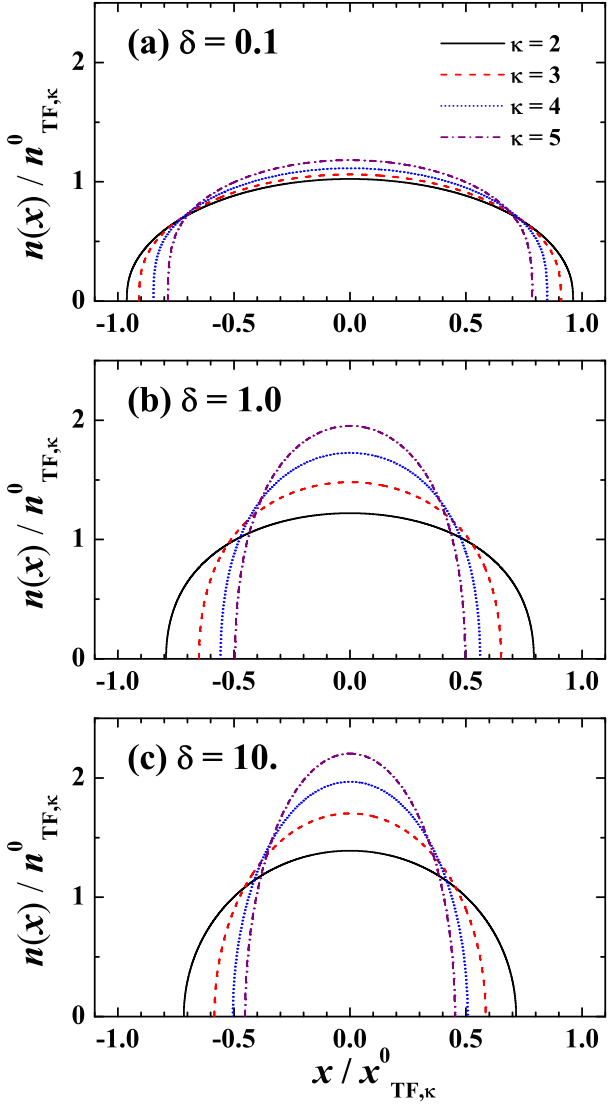} 
\par\end{centering}

\caption{(Color online) Density profiles of a 1D trapped multi-component Fermi
cloud at three interaction parameters $\delta=0.1$ (a), $\delta=1.0$
(b), and $\delta=10$ (c). The linear density and the coordinate are
in units of the peak density $n_{TF,\kappa}^{0}$ and Thomas-Fermi
radius $x_{TF,\kappa}^{0}$ of an ideal gas, respectively.}

\label{fig5} 
\end{figure}

%
\begin{figure}
\begin{centering}
\includegraphics[clip,width=0.45\textwidth]{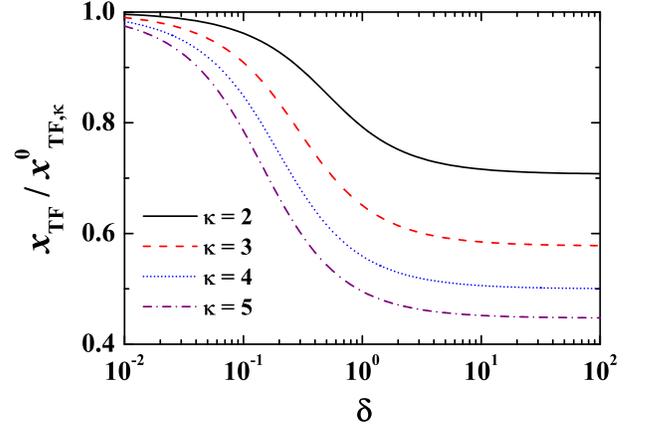} 
\par\end{centering}

\caption{(Color online) Thomas-Fermi radius, in units of $x_{TF,\kappa}^{0}$,
as a function of the interaction parameter, at several number of component
as indicated.}

\label{fig6} 
\end{figure}

%
\begin{figure}
\begin{centering}
\includegraphics[clip,width=0.45\textwidth]{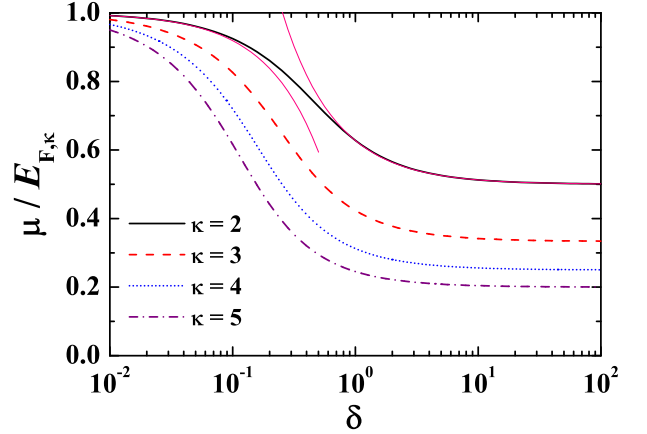} 
\par\end{centering}

\caption{(Color online) Chemical potential as a function of the interaction
parameter $\delta$. It is in units of $E_{F,\kappa}=N\hbar\omega/\kappa$.
Thin solid lines are the analytic results in the two limiting cases
for $\kappa=2$, as described in Eqs. (\ref{muTG}) and (\ref{muID}).}

\label{fig7} 
\end{figure}

Accordingly, the chemical potential takes the form, \begin{equation}
\mu_{g}^{\delta\gg1}=-\frac{\left(\kappa^{2}-1\right)\hbar^{2}}{6ma_{1D}^{2}}+\frac{N\hbar\omega}{\kappa^{2}}\left[1+\frac{64\sqrt{2}\kappa S_{\kappa}}{9\pi^{2}}\frac{1}{\delta}\right].\label{muTG}\end{equation}
 where the first term on the right hand side is again from the binding
energy, while the second term corresponds to the chemical potential
of a spinless Tonks-Girardeau gas of bound clusters. For later reference,
we calculate the mean-square size of the cloud $\left\langle x^{2}\right\rangle =\int dxx^{2}n(x)/N$
using the strongly interacting density profile $n\left(x\right)_{\delta\gg1}$,
and find that, \begin{equation}
\left\langle x^{2}\right\rangle _{\delta\gg1}=\frac{N}{2\kappa^{2}}a_{ho}^{2}+\frac{32\sqrt{2}S_{\kappa}}{15\pi^{2}\kappa}N^{3/2}a_{1D}a_{ho}.\label{x2TG}\end{equation}

The density profile of a weakly interacting gas can be calculated
in the same manner. The leading order contribution is simply the ideal
gas result, $n_{ideal}(x)$, and the Hartree-Fock correction to the
chemical potential gives rise to a density variation which is proportional
to the interaction strength $g_{1D}$. Explicitly, we find that, \begin{equation}
n\left(x\right)_{\delta\ll1}=n_{ideal}\left(x\right)+\Delta n\left(x\right),\end{equation}
 where \begin{equation}
\Delta n\left(x\right)=\frac{2\kappa\left(\kappa-1\right)}{\pi^{2}a_{1D}}\left[1-\frac{2/\pi}{\sqrt{1-x^{2}/\left(x_{TF,\kappa}^{0}\right)^{2}}}\right].\end{equation}
 The chemical potential is given by, \begin{equation}
\mu_{g}^{\delta\ll1}=\frac{N\hbar\omega}{\kappa}\left[1-\frac{4\sqrt{2\kappa}\left(\kappa-1\right)}{\pi^{2}}\delta\right].\label{muID}\end{equation}
 The mean-square size of the cloud is found to be, \begin{equation}
\left\langle x^{2}\right\rangle _{\delta\ll1}=\frac{N}{2\kappa}a_{ho}^{2}-\frac{4}{3}\sqrt{\frac{2}{\kappa}}\frac{\kappa-1}{\pi^{2}}N^{1/2}\frac{a_{ho}^{3}}{a_{1D}}.\label{x2ID}\end{equation}

Fig. (\ref{fig5}) plots numerical results for the density profiles
at three interaction parameters and at different number of species
as indicated. The linear density and the coordinate are in units of
the peak density $n_{TF,\kappa}^{0}$ and Thomas-Fermi radius $x_{TF,\kappa}^{0}$
of an ideal gas, respectively. With increasing interaction parameter
$\delta$ (from Figs. (\ref{fig5}a) to (\ref{fig5}c)), the density
profiles change from an ideal gas distribution to a strongly interacting
Tonks-Girardeau profile. At the same interaction parameter, the density
profiles become sharper and narrower as the number of species $\kappa$
increases. We display also the Thomas-Fermi radius of the cloud in
Fig. (\ref{fig6}) as a function of the interaction strength. It decreases
monotonically from the ideal gas result $x_{TF,\kappa}^{0}$ to the
Tonks-Girardeau prediction $x_{TF,\kappa}^{0}/\sqrt{\kappa}$ as the
interaction parameter increases, thus increasing the compressive force
of the attractive interactions.

Fig. (\ref{fig7}) reports the dependence of the chemical potential
on the interaction strength, at several numbers of species as indicated.
Here, for clarity we have subtracted the binding energy part of the
chemical potential. For $\kappa=2$, we show the asymptotic behavior
given by Eqs. (\ref{muTG}) and (\ref{muID}) by thin solid lines.
They fit very well with the numerical results, except at the intermediate
interaction regime.

\subsection{Low-lying collective modes}

Experimentally, a useful way to characterize an interacting system
is to measure its low-lying collective excitations of density oscillations.
For a uniform gas, the low-lying collective excitations are simply
the sound waves with energy $\Omega(k)=c\left|{\bf k}\right|$ for
a given momentum $k<k_{F}$, which is gapless as $k\rightarrow0$.
In traps, however, the spectrum becomes discrete, due to the finite
cloud size of the gas that is of order $x_{TF,\kappa}^{0}$. There
is a minimum value of the momentum $k_{\min}\sim1/x_{TF,\kappa}^{0}$.
Taking a sound velocity at the trap center $c\sim(\hbar n_{TF,\kappa}^{0}/m)(\pi/\kappa)$,
one finds that, \begin{equation}
\Omega(k_{\min})\sim\frac{\hbar n_{TF,\kappa}^{0}}{m}\frac{\pi}{\kappa}\frac{1}{x_{TF,\kappa}^{0}}=\omega,\end{equation}
 comparable to the energy level of the harmonic trap.

Since the charge degree of freedom of the gas falls into the Luttinger
liquid universality class, quantitative calculations of the low-lying
collective excitations in traps can be carried out based on the superfluid
hydrodynamic description of the dynamics of the 1D Fermi gas \citep{stringari,ldh1d}.
In such a description, the density $n\left(x,t\right)$ and the velocity
field $v\left(x,t\right)$ satisfy the equation of continuity \begin{equation}
\frac{\partial n\left(x,t\right)}{\partial t}+\frac{\partial}{\partial x}\left[n\left(x,t\right)v\left(x,t\right)\right]=0,\end{equation}
 and the Euler equation \begin{equation}
m\frac{\partial v}{\partial t}{\bf +}\frac{\partial}{\partial x}\left[\mu_{\hom}\left(n\right)+V_{trap}\left(x\right)+\frac{1}{2}mv^{2}\right]=0.\end{equation}
 We consider the fluctuations of the density and the velocity field
about the equilibrium ground state , $\delta n\left(x,t\right)=$
$n\left(x,t\right)-n(x)$ and $\delta v\left(x,t\right)=v\left(x,t\right)-v(x)=v\left(x,t\right)$,
where $n(x)$ and $v(x)\equiv0$ are the equilibrium density profile
and velocity field. Linearizing the hydrodynamic equations, one finds
that \citep{stringari}, \begin{equation}
\frac{\partial^{2}}{\partial t^{2}}\delta n\left(x,t\right)=\frac{1}{m}\frac{\partial}{\partial x}\left\{ n\frac{\partial}{\partial x}\left[\frac{\partial\mu_{\hom}(n)}{\partial n}\delta n\left(x,t\right)\right]\right\} .\end{equation}
 The boundary condition requires that the current $j(x,t)=n(x)\delta v\left(x,t\right)$
should vanish identically at the Thomas-Fermi radius $x=\pm x_{TF}$.
Considering the $n$th eigenmode with $\delta n\left(x,t\right)=\delta n\left(x\right)\exp\left[i\omega_{n}t\right]$
and removing the time-dependence, we end up with an eigenvalue problem,
\textit{i.e.}, \begin{equation}
\frac{1}{m}\frac{d}{dx}\left\{ n\frac{d}{dx}\left[\frac{\partial\mu_{\hom}(n)}{\partial n}\delta n\left(x\right)\right]\right\} +\omega_{n}^{2}\delta n\left(x\right)=0.\end{equation}

We develop a powerful multi-series-expansion method to solve the above
1D hydrodynamics equation, as outlined in detail in the Appendix.
The resulting low-lying collective mode can be classified by the number
of nodes in its eigenfunction, \textit{i.e.}, the number index {}``$n$''.
The lowest two modes with $n=1,2$ have very transparent physical
meaning. These are respectively the dipole and breathing (compressional)
modes, which can be excited separately by shifting the trap center
or modulating the harmonic trapping frequency. The dipole mode is
not affected by interactions according to Kohn's theorem, and has
an invariant frequency precisely at $\omega_{1}=\omega$.

The low-lying hydrodynamic modes of a two-component Fermi gas in the
weak and strong coupling limits have been discussed analytically by
Minguzzi \citep{anna}. In both limits, the cloud behaves like a spinless
ideal Fermi gas. Therefore, the frequency $\omega_{n}$ of the mode
{}``$n$'' is fixed to $n\omega$. This result applies to a multi-component
1D Fermi gas as well.

%
\begin{figure}
\begin{centering}
\includegraphics[clip,width=0.45\textwidth]{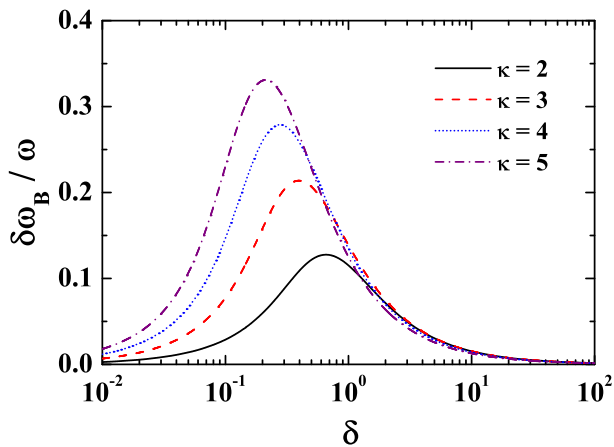} 
\par\end{centering}

\caption{(Color online) Frequency corrections (with respect to the ideal gas
value) of the lowest breathing modes $\delta\omega_{B}$ as a function
of the dimensionless coupling parameter $\delta$, at several number
of species as indicated.}

\label{fig8} 
\end{figure}

Fig. (\ref{fig8}) shows the interaction dependence of frequencies
of the breathing mode for a multi-component gas with different numbers
of components. Here, to stress the role of interactions, the deviation
of the mode frequency from its ideal result, $\delta\omega_{B}=$
$\omega_{B}-2\omega$, has been plotted. As a function of the interaction
parameter, a peak in $\delta\omega_{B}$ emerges at the intermediate
interaction regime $\delta\sim1$. The peak value increases with increasing
the number of components $\kappa$. This peak is a clear signature
of the cross-over from a regime of multi-particle clusters to a color
quasi-superconductor.

%
\begin{figure}
\begin{centering}
\includegraphics[clip,width=0.45\textwidth]{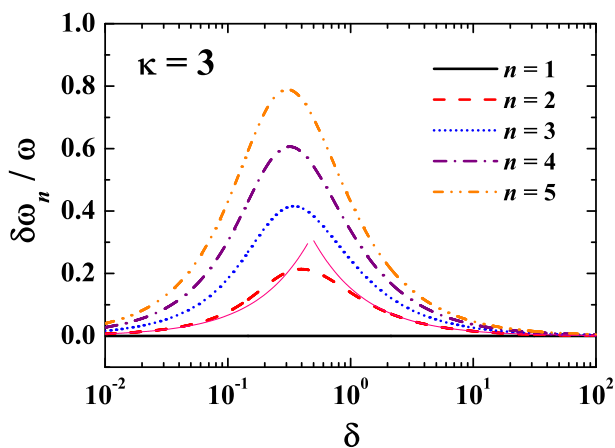} 
\par\end{centering}

\caption{(Color online) Interaction strength dependence of the frequency corrections
$\delta\omega_{n}=\omega_{n}-n\omega$ of the low-lying collective
modes of a 1D three-component Fermi gas. Thin solid lines are the
analytic sum-rule expansions for the weakly and strongly interacting
limits, as described in Eqs. (\ref{dwTG}) and (\ref{dwID}), respectively.}

\label{fig9} 
\end{figure}

Fig. (\ref{fig9}) shows the frequency correction $\delta\omega_{n}=$
$\omega_{n}-n\omega$ of the low-lying modes of a three-component
gas as a function of the interaction strength. The dipole mode frequency
is always $\omega$, as expected, while all other modes show a non-trivial
peak structure, similar to that of the breathing mode. It is evident
that the higher mode index {}``$n$'' is, the larger frequency correction.

As an alternative to the numerical solution of the 1D hydrodynamic
equation, the frequency of the breathing mode may be obtained by applying
the compressibility sum-rule. As long as the breathing mode exhausts
all the weights in the dynamic structure factor, the mode frequency
can be calculated according to \citep{astrakharchik,stringari}, \begin{equation}
\omega_{B}^{2}=-2\frac{\left\langle x^{2}\right\rangle }{d\left\langle x^{2}\right\rangle /d\omega^{2}}.\label{sumrules}\end{equation}
 We have performed such calculations. The resulting frequency agrees
extremely well with that from solving the 1D hydrodynamic equation,
with a relative difference less than $10^{-4}$. Such a good agreement
gives strong support to the application of the sum-rule. To gain further
insight of the breathing mode frequency, we use the sum-rule to obtain
analytically the first order correction to the frequency for the weak
and strong coupling limits. This can be done by substituting the expression
for the mean-square size of the cloud in Eqs. (\ref{x2TG}) and (\ref{x2ID})
into the sum-rule formalism (\ref{sumrules}). Replacing $a_{ho}=\sqrt{\hbar/m\omega}$
and taking the derivative with respect to $\omega^{2}$, we find that
\begin{equation}
\omega_{B}^{\delta\gg1}=2\omega\left(1+\frac{16\sqrt{2}\kappa S_{\kappa}}{15\pi^{2}}\frac{1}{\delta}\right),\label{dwTG}\end{equation}
 for the strongly interacting limit, and \begin{equation}
\omega_{B}^{\delta\ll1}=2\omega\left[1+\frac{2\sqrt{2\kappa}\left(\kappa-1\right)}{3\pi^{2}}\delta\right],\label{dwID}\end{equation}
 for the weak coupling regime. Thin solid lines in Fig. (\ref{fig9})
show the resulting analytic expansions for $\delta\omega_{B}$ for
a three-component Fermi gas. These expansions describe very well the
breathing mode frequency over a fairly large range of interaction
strengths, but break down at the intermediate coupling regime $\delta\sim1$.

\section{Conclusion}

In conclusion, we have investigated the properties of 1D attractive
multi-component Fermi gases, based on the Bethe ansatz exact solution,
in a homogeneous environment. This was extended to include a harmonic
trap, by using the local density approximation. The equation of state
of the system has been discussed in detail, as well as some dynamical
quantities, including the sound velocity and low-lying collective
modes.

We have drawn attention to the formation of multi-body bound clusters,
which are always present in a 1D attractive multi-component gas. These
clusters are not well understood, as their description is beyond the
mean-field theory. We have found that such multi-body clusters have
significant impact on the equation of state and dynamic behavior of
the systems. In particular, as the number of particles in the clusters
increase, the gas turns out to be increasingly attractive compared
to a simple Cooper-pairing scenario. This is suggestive of the strongly
interacting nature of the bound clusters.

There is a close analogy between the proposed properties of QCD at
high density, and the results we calculate for a three-component Fermi
gas in 1D. In both cases there is a transition between discrete multi-particle
clusters (nucleons) to a coherent quasi-superfluid (colour superconductor),
as the density increases. For the 1D case, both the high and low density
limits result in free particle dynamics, with different multiplicities.
This gives simple collective mode behaviour in either limit. However,
the transition region with $\gamma\simeq1$ can show very strong evidence
of a transition. This is due to an easily observed peak in the collective
breathing mode frequency.

Our results should be useful for the future experiments on 1D multi-component
atomic Fermi gases. An example of particular interest is three-component
lithium gas, which has three broad Feshbach resonances. Our estimate
of the relevant parameters suggests that an ultra-cold gas of $^{6}$Li
atoms in an optical lattice, above a magnetic field $B=1000$ Gauss,
can be nearly described by the current model. Our prediction of the
density profiles and low-lying collective modes, provide a useful
characterization of an interacting 1D three-component Fermi gas. We
expect this to be testable in a future experiment.


\begin{acknowledgments}
This work was supported by the Australian Research Council Center
of Excellence, the National Natural Science Foundation of China Grants
Nos. NSFC-10574080 and NSFC-10774190, and the National Fundamental
Research Program of China Grants Nos. 2006CB921404 and 2006CB921306. 
\end{acknowledgments}

\section*{Appendix}

In this appendix, we outline the procedure of solving the 1D hydrodynamic
equation, \begin{equation}
\frac{1}{m}\frac{d}{dx}\left\{ n\frac{d}{dx}\left[\frac{\partial\mu_{\hom}(n)}{\partial n}\delta n\left(x\right)\right]\right\} +\omega_{n}^{2}\delta n\left(x\right)=0,\end{equation}
 with the boundary condition that the current $j(x)\propto\partial_{x}\{\delta n\left(x\right)\partial\mu_{\hom}[n(x)]/\partial n(x)\}$
must vanish at $x=\pm x_{TF}$.

It is convenient to introduce a function, \begin{equation}
f\left(x\right)\equiv\frac{\partial\mu_{\hom}\left[n\left(x\right)\right]}{\partial n\left(x\right)}\delta n\left(x\right),\end{equation}
 and rewrite the hydrodynamic equation in the form, \begin{equation}
x\frac{d^{2}f}{dx^{2}}+x\frac{d\ln\left[n(x)\right]}{dx}\frac{df}{dx}+\left(\frac{\omega_{m}}{\omega}\right)^{2}\frac{\omega^{2}x}{c^{2}(x)}f=0,\end{equation}
 where $c(x)=[n(x)\partial\mu_{\hom}[n(x)]/\partial n(x)/m]^{1/2}$
is the local sound velocity at position $x$. Let us introduce $A\left(x\right)\equiv\ln\left[n(x)\right]$,\ $A_{1}\left(x\right)\equiv-\omega^{2}x/c^{2}\left(x\right)$,
and $\nu_{m}\equiv\omega_{m}/\omega$. Then the above equation becomes,
\begin{equation}
x\frac{d^{2}f}{dx^{2}}+x\frac{dA}{dx}\frac{df}{dx}-\nu_{m}^{2}A_{1}\left(x\right)f=0.\end{equation}
 Here the variable $x\in\left[-x_{TF},+x_{TF}\right]$. Note that
for our specific case, $A_{1}\left(x\right)=dA(x)/dx$, due to the
local equilibrium condition.

To solve the equation, one may wish to change $x$ to a new variable
$y\in\left[0,1\right]$. This can be done by two steps. First, we
define $f\left(x\right)\equiv x^{l}\upsilon\left(x\right)$, where
$l=0$ or $1$ corresponds to the parity of the modes. Translating
to $\upsilon\left(x\right)$, we have (now $x\in\left[0,+x_{TF}\right]$),
\begin{equation}
x\frac{d^{2}\upsilon}{dx^{2}}+\left(2l+x\frac{dA}{dx}\right)\frac{d\upsilon}{dx}-\left[\nu_{m}^{2}A_{1}-l\frac{dA}{dx}\right]\upsilon=0.\end{equation}
 At the second step, we define $y\equiv\left(x/x_{TF}\right)^{2}$
and change to the new variable $y$. After some straightforward calculations,
we obtain the following equation for $\upsilon\left(y\right)$, \begin{eqnarray}
 &  & y\left(1-y\right)^{2}\frac{d^{2}\upsilon}{dy^{2}}+\left(1-y\right)^{2}\left[\Delta+y\frac{dA}{dy}\right]\frac{d\upsilon}{dy}\nonumber \\
 &  & \left.+\frac{1}{2}\left(1-y\right)^{2}\left[l\frac{dA}{dy}-\nu_{m}^{2}\tilde{A}_{1}\left(y\right)\right]\upsilon=0\right.,\label{hydroeq}\end{eqnarray}
 where $\Delta\equiv l+1/2$ and $\tilde{A}_{1}\left(y\right)\equiv A_{1}(y)/\left(2y^{1/2}\right)$.
In practise, the value of $dA/dy$ can be calculated as follows, \begin{equation}
\frac{dA}{dy}\equiv-\frac{m\omega^{2}x_{TF}^{2}}{2}\left[\frac{1}{n\partial\mu_{\hom}[n]/\partial n}\right]_{n=n(x=x_{TF}\sqrt{y})}.\end{equation}
 Note that in our case $\tilde{A}_{1}\left(y\right)=dA/dy$. Here,
we multiply a factor of $\left(1-y\right)^{2}$ on both sides of Eq.
(\ref{hydroeq}), in order to remove the singularity of $dA/dy$ and
$\tilde{A}_{1}\left(y\right)$\ at point $y=1$.

We develop a multi-series-expansion method to solve the eigenvalue
problem Eq. (\ref{hydroeq}). As the current vanishes at the Thomas-Fermi
boundary, the eigenfunction of Eq. (\ref{hydroeq}) should not be
singular at $y=1$. As well, we require that the eigenfunction has
to take a finite value at $y=0$. As we shall see, these two boundary
conditions give rise to a set of the discrete spectrum, as well as
a complete set of the eigenfunctions.

To apply the boundary conditions, we divide the whole region $\left[0,1\right]$
into many pieces, for example, $M$ parts, $\left[0,1\right]=\left[y_{0}=0,y_{1}\right]\cup\left[y_{1},y_{2}\right]\cup...\cup\left[y_{M-1},y_{M}=1\right]$.
We look for the solution in the form, \begin{equation}
\upsilon\left(y\right)=\sum_{n=0}^{\infty}a_{in}\left(y-y_{i}\right)^{n},\quad\text{if{\rm \ }}y\subset\left[y_{i},y_{i+1}\right].\end{equation}
 Hence, the two boundary conditions translate to the requirement of
a well-convergent series of $\left\{ a_{in}\right\} $ at both the
starting region $\left[y_{0}=0,y_{1}\right]$ and the ending region
$\left[y_{M-1},y_{M}=1\right]$. The basic idea of solving the eigenvalue
problem is then clear. We use the strategy of try and test. Given
the parity $l$, we make an initial guess for $\nu_{m}$, and setup
the series $\left\{ a_{in}\right\} $ at the starting region $\left[y_{0}=0,y_{1}\right]$,
and then propagate it to the ending region of $\left[y_{M-1},y_{M}=1\right]$.
If the series converges at $y=1$, then we find a correct eigenvalue
and eigenfunction of the problem. Otherwise, we scan the value of
$\nu_{m}$, until all the required low-lying eigenvalues are found.

In greater detail, we apply the try and test strategy as follows.
(\textbf{A}) At first, let us approximate, at each region $\left[y_{i},y_{i+1}\right]$,
\begin{eqnarray}
-(1-y)^{2}\frac{dA}{dy} & = & \tilde{p}_{0}+\tilde{p}_{1}y+\tilde{p}_{2}y^{2},\\
-(1-y)^{2}\tilde{A}_{1}\left(y\right) & = & \tilde{q}_{0}+\tilde{q}_{1}y+\tilde{q}_{2}y^{2}.\end{eqnarray}
 To make the expansion accurate, generally we take $M\sim30$. By
introducing a new variable $z=y-y_{i}$, at the region $\left[y_{i},y_{i+1}\right]$
we can cast the Eq. (\ref{hydroeq}) into the form, \begin{equation}
\left(\sum_{j=0}^{3}r_{j}z^{j}\right)\frac{d^{2}\upsilon}{dy^{2}}+\left(\sum_{j=0}^{3}p_{j}z^{j}\right)\frac{d\upsilon}{dy}+\left(\sum_{j=0}^{3}q_{j}z^{j}\right)\upsilon=0,\end{equation}
 where the coefficients $\left\{ r_{i}\right\} ,$ $\left\{ p_{i}\right\} $
and $\left\{ q_{i}\right\} $ can be calculated directly from $\left\{ \tilde{p}_{i}\right\} $
and $\left\{ \tilde{q}_{i}\right\} $ in the program. We then substitute
the solution $\upsilon\left(z\right)=\sum_{n=0}^{\infty}a_{in}z^{n}$\ into
the above equation to obtain the iterative relation (without confusion,
here we denote $a_{n}\equiv a_{in}$ for this region), \begin{eqnarray}
a_{n+2} & = & -\frac{\left(n+1\right)\left(nr_{1}+p_{0}\right)}{\left(n+2\right)\left(n+1\right)r_{0}}a_{n+1}\nonumber \\
 &  & -\frac{\left[n\left(n-1\right)r_{2}+np_{1}+q_{0}\right]}{\left(n+2\right)\left(n+1\right)r_{0}}a_{n}\nonumber \\
 &  & -\frac{\left[\left(n-1\right)\left(n-2\right)r_{3}+\left(n-1\right)p_{2}+q_{1}\right]}{\left(n+2\right)\left(n+1\right)r_{0}}a_{n-1}\nonumber \\
 &  & -\frac{\left[\left(n-2\right)p_{3}+q_{2}\right]}{\left(n+2\right)\left(n+1\right)r_{0}}.\label{series}\end{eqnarray}
 We need to classify two cases. (i) In the starting region of $y_{0}=0$,
we have $r_{0}=0$ due to the boundary condition. Up to an overall
irrelevant factor, we can set $a_{0}=1$\ and then $a_{1}=-q_{0}/p_{0}$.
(ii) In the other regions, $a_{0}$\ and $a_{1}$\ are determined
by the continuous conditions as stated below. Once $a_{0}$\ and
$a_{1}$ are known, we could obtain all the values of $a_{n}$ by
Eq. (\ref{series}) since $a_{-1}=a_{-2}=0$. Usually it is already
sufficiently accurate to keep $n\leq n_{\max}=16$. (\textbf{B}) The
series $\left\{ a_{in}\right\} $ at different regions are connected
by the requirement that the function $\upsilon\left(y\right)$\ and
its first derivation $\upsilon^{\prime}\left(y\right)$ should be
continuous at the point $\left\{ y_{i}\right\} $, where the index
$i$ runs from $1$\ to $M-1$. (\textbf{C}) In this way, we can
finally obtain the series $\left\{ a_{in}\right\} $\ at the region
$\left[y_{M-1},y_{M}\right]$. We judge the convergence by checking
whether the value of $a_{M,n_{\max}}$\ is sufficiently small or
not.

In practise, the above procedure of solving the 1D hydrodynamic equation
of a multi-component Fermi gas is very efficient and accurate. It
can be applied as well to other 1D systems, such as the 1D interacting
Bose gas, and to the 3D systems in spherical harmonic traps.


\begin{thebibliography}{10}
\bibitem{FR} S. Inouye \textit{et al.}, Nature (London) \textbf{392},
151 (1998).

\bibitem{lattice} M. Greiner \textit{et al.}, Nature (London) \textbf{415},
39 (2002).

\bibitem{huinp} H. Hu, P. D. Drummond, and X.-J. Liu, Nature Physics
\textbf{3}, 469 (2007), and references therein.

\bibitem{leggett} A. J. Leggett, \textit{Modern Trends in the Theory
of Condensed Matter} (Springer-Verlag, Berlin, 1980).

\bibitem{nsr} P. Nozières, and S. Schmitt-Rink, J. Low Temp. Phys.
\textbf{59}, 195 (1985).

\bibitem{randeria} J. R. Engelbrecht, M. Randeria, and C. A. R. S?de Melo, Phys. Rev. B \textbf{55}, 15153 (1997).

\bibitem{griffin} Y. Ohashi and A. Griffin, Phys. Rev. Lett. \textbf{89},
130402 (2002).

\bibitem{haussmann} R. Haussmann \textit{et al.}, Phys. Rev. A \textbf{75},
023610 (2007).

\bibitem{hui04} H. Hu \textit{et al.}, Phys. Rev. Lett. \textbf{93},
190403 (2004).

\bibitem{hld} H. Hu, X.-J. Liu, and P. D. Drummond, Europhys. Lett.
\textbf{74}, 574 (2006).

\bibitem{jila} C. A. Regal and D. S. Jin, Phys. Rev. Lett. \textbf{92},
040403 (2004).

\bibitem{mit04} M. W. Zwierlein \textit{et al.}, Phys. Rev. Lett.
\textbf{92}, 120403 (2004).

\bibitem{duke} J. Kinast \textit{et al.}, Phys. Rev. Lett. \textbf{92},
150402 (2004).

\bibitem{chin} C. Chin \textit{et al.}, Science \textbf{305}, 1128
(2004).

\bibitem{ens} T. Bourdel \textit{et al.}, Phys. Rev. Lett. \textbf{93},
050401 (2004).

\bibitem{mit05} M. W. Zwierlein \textit{et al.}, Nature (London)
\textbf{435}, 1047 (2005).

\bibitem{randy05} G. B. Partridge \textit{et al.}, Phys. Rev. Lett.
\textbf{95}, 020404 (2005).

\bibitem{randy06} G. B. Partridge \textit{et al.}, Science \textbf{311},
503 (2006).

\bibitem{luu} T. Luu and A. Schwenk, Phys. Rev. Lett. \textbf{98},
103202 (2007).

\bibitem{efimov} V. Efimov, Nucl. Phys. A \textbf{210}, 157 (1973).

\bibitem{grimm} T. Kraemer \textit{et al.}, Nature \textbf{440},
315(2006).

\bibitem{hofstetter3} Ákos Rapp, G. Zaránd, C. Honerkamp, and W.
Hofstetter, Phys. Rev. Lett. \textbf{98}, 160405 (2007).

\bibitem{BailinLove84}D. Bailin and A. Love, Phys. Rep. \textbf{107},
325 (1984).

\bibitem{Alford}M. G. Alford, K. Rajagopal, F. Wilczek, Phys. Lett.
B \textbf{422}, 247(1998); M. A. Halasz, A. D. Jackson, R. E. Shrock,
M. A. Stephanov, and J. J. M. Verbaarschot, Phys. Rev. D \textbf{58},
096007 (1998), N. Itoh, Prog. Theor. Phys. \textbf{44}, 291 (1970).

\bibitem{Allton}C. R. Allton, S. Ejiri, S. J. Hands, O. Kaczmarek,
F. Karsch, E. Laermann, Ch. Schmidt, and L. Scorzato, Phys. Rev. D
\textbf{ 66}, 074507 (2002).

\bibitem{esslinger} H. Moritz, T. Stöferle, M. Köhl, and T. Esslinger,
Phys. Rev. Lett. \textbf{91}, 250402 (2003).

\bibitem{a3dB} M. Bartenstein \textit{et al.}, Phys. Rev. Lett. \textbf{94},
103201 (2005).

\bibitem{hofstetter1} C. Honerkamp and W. Hofstetter, Phys. Rev.
Lett. \textbf{92}, 170403 (2004).

\bibitem{hofstetter2} C. Honerkamp and W. Hofstetter, Phys. Rev.
B \textbf{70}, 094521 (2004).

\bibitem{torma1} T. Paananen, J.-P. Martikainen, and P. Törm? Phys.
Rev. A \textbf{73}, 053606 (2006).

\bibitem{lianyi} L. He, M. Jin, and P. Zhuang, Phys. Rev. A \textbf{74},
033604 (2006).

\bibitem{cherng} R. W. Cherng, G. Refael, and E. Demler, Phys. Rev.
Lett. \textbf{99}, 130406 (2007).

\bibitem{bedaque} P. F. Bedaque and J. P. D'Incao, arXiv:cond-mat/0602525.

\bibitem{zhai} H. Zhai, Phys. Rev. A \textbf{75}, 031603(R) (2007).

\bibitem{sedrakian} A. Sedrakian and J. W. Clark, Phys. Rev. C \textbf{73},
035803 (2006).

\bibitem{torma2} T. Paananen, P. Törm? and J.-P. Martikainen, Phys.
Rev. A \textbf{75}, 023622 (2007).

\bibitem{Capponi}S. Capponi, G. Roux, P. Lecheminant, P. Azaria,
E. Boulat, and S. R. White, arXiv:0706.0609v1.

\bibitem{Lecheminant} P. Lecheminant, E. Boulat, and P. Azaria, Phys.
Rev. Lett. \textbf{95}, 240402 (2005).

\bibitem{Wu} Congjun Wu, Phys. Rev. Lett. \textbf{95}, 266404 (2005).

\bibitem{lieb} E. H. Lieb and W. Liniger, Phys. Rev. \textbf{130},
1605 (1963).

\bibitem{yang} C. N. Yang, Phys. Rev. Lett. \textbf{19}, 1312 (1967).

\bibitem{gaudin} M. Gaudin, Phys. Lett. \textbf{24A}, 55(1967).

\bibitem{takahashi1} T. Takahashi, \textit{Thermodynamics of One-Dimensional
Solvable Models} (Cambridge University Press, Cambridge, 1999).

\bibitem{bergeman} T. Bergeman, M. G. Moore, and M. Olshanii, Phys.
Rev. Lett. \textbf{91}, 163201 (2003).

\bibitem{astrakharchik} G. E. Astrakharchik, D. Blume, S. Giorgini,
and L. P. Pitaevskii, Phys. Rev. Lett. \textbf{93}, 050402 (2004).

\bibitem{footnote} Note the difference in the definition of $a_{\rho}$
with \citep{bergeman}, which accounts for $A=-\zeta(1/2)/\sqrt{2}\simeq1.0326$.

\bibitem{McGuire}I. B. McGuire, J. Math. Phys. \textbf{5}, 622 (1964);
H. B. Thacker, Rev. Mod. Phys. \textbf{53}, 253 (1981).

\bibitem{DrummondNature} P. D. Drummond, R. M. Shelby, S. R. Friberg
and Y. Yamamoto, Nature \textbf{365}, 307 (1993).

\bibitem{Strecker}K. S. Strecker, G. B. Partridge, A. G. Truscott,
and R. G. Hulet, Nature \textbf{417}, 150 (2002); L. Khaykovich, F.
Schreck, G. Ferrari, T. Bourdel, J. Cubizolles, L. D. Carr, Y. Castin,
C. Salomon, Science \textbf{296}, 1290 (2002); Simon L. Cornish, Sarah
T. Thompson, and Carl E. Wieman, Phys. Rev. Lett. \textbf{96}, 170401
(2006).

\bibitem{zwerger} W. Zwerger, J. Opt. B: Quantum Semiclass. Opt.
\textbf{5}, S9 (2003).

\bibitem{ldh1d} X.-J. Liu, P. D. Drummond, and H. Hu, Phys. Rev.
Lett. \textbf{94}, 136406 (2005).

\bibitem{hld1d} H. Hu, X.-J. Liu, and P. D. Drummond, Phys. Rev.
Lett. \textbf{98}, 070403 (2007); X.-J. Liu, H. Hu, and P. D. Drummond,
Phys. Rev. A \textbf{76}, 043605 (2007).

\bibitem{guan1} M. T. Batchelor \textit{et al.,} Journal of Physics
Conference Series \textbf{42}, 5 (2006).

\bibitem{guan2} X.-W. Guan \textit{et al.}, Phys. Rev. B \textbf{76},
085120 (2007).

\bibitem{guan3} X.-W. Guan \textit{et al.}, arXiv: 0709.1763.

\bibitem{takahashi2} M. Takahashi, Prog. Theor. Phys. \textbf{44},
899 (1970).

\bibitem{Schlottmann}P. Schlottmann, J.Phys.Condens. Matter.\textbf{6},1359
(1994).

\bibitem{tg} M. D. Girardeau, J. Math. Phys. \textbf{1}, 516 (1960).

\bibitem{heiselberg} H. Heiselberg, Phys. Rev. A \textbf{63}, 043606
(2001).

\bibitem{stringari} C. Menotti and S. Stringari, Phys. Rev. A \textbf{66},
043610 (2002).

\bibitem{anna} A. Minguzzi, P. Vignolo, M. L. Chiofalo, and M. P.
Tosi, Phys. Rev. A \textbf{64}, 033605 (2001). 
\end{thebibliography}
\end{document}